\documentclass{iaus}
\usepackage{graphicx}

\title[Abundances on the Main Sequence of $\omega$ Cen] 
{Abundances on the Main Sequence of the Globular Cluster $\omega$ Centauri}

\author[Da Costa, Stanford, Norris \& Cannon]   
{G. S. Da Costa$^1$, 
Laura M. Stanford$^1$, John E. Norris$^1$ \break \and R. D. Cannon$^2$}

\affiliation{$^1$Research School of Astronomy \& Astrophysics, The Australian 
National Observatory \break 
email: gdc, stanford, jen@mso.anu.edu.au\\[\affilskip]
$^2$Anglo-Australian Observatory \break email: rdc@aaoepp.aao.gov.au}

\pubyear{2005}
\volume{228}  
\pagerange{1--7}
\date{\today and in revised form ??}
\jname{From Lithium to Uranium: Elemental Tracers of Early Cosmic Evolution}
\editors{V. Hill, P. Fran\c{c}ois \& F. Primas, eds.}
\begin{document}

\maketitle

\begin{abstract}
Using the 2dF multi-fibre instrument on the Anglo-Australian Telescope,
moderate resolution spectra have been obtained for a large sample of
stars on the main sequence and at the turnoff in the unusual globular
cluster $\omega$ Cen.  We investigate the 
behaviour of CH, CN and SrII line strength indices as a function of overall
abundance for the main sequence sample.  A number of stars do not follow
 the relations defined by the majority.  These
anomalous objects can be categorized into (at least) three types. (1) Carbon 
enhanced stars, which represent about 5\% of the sample, and which are
found at all metallicities.  Spectrum synthesis
calculations show that the atmospheres of these stars are typically 
enhanced in carbon by
factors of between 3 and 10.  (2) Nitrogen enhanced stars, revealed for
[Fe/H] $\geq$ --1.3 by strong CN indices, which make up $\sim$40\% of 
the cluster main sequence population above this metallicity.  The
stars are enhanced in nitrogen by factors of up to 100.  Our data, however,
provide no constraints on their relative numbers at lower [Fe/H].   
 (3) Stars with enhancements of the 
s-process element Sr by factors of 30 to 60.  The possible origins for these
abundance anomalies are discussed.  
\keywords{globular clusters: individual ($\omega$ Cen), stars: abundances,
stars: AGB and post-AGB, stars: chemically peculiar}
\end{abstract}

\firstsection 
\section{Introduction}

The globular cluster $\omega$ Centauri has been recognised as an 
unusual object for at least three decades.  Foremost among its peculiar
characteristics is the large abundance range exhibited by the cluster
stars: from [Fe/H] $\approx$ --1.8 to least [Fe/H] 
$\approx$ --0.5 (e.g.\  \cite[Pancino et al.\ 2002, and the references
therein]{EP02}).  Such a 
large abundance range is not observed in any other globular cluster.
There is also evidence that the star formation in $\omega$~Cen took place
over an extended period.  This contention
is based on both spectroscopic and photometric analyses.  For example,
high dispersion studies (e.g.\ \cite[Norris \& Da~Costa 1995b, 
Smith et al.\ 2000]{ND95b,VS00}) have revealed the
chemical enrichment signatures of low mass ($\sim$1.5--3~M$_{\odot}$) AGB 
stars in the cluster. Such stars have lifetimes of 1--3 Gyr.
Similarly, population modelling of the horizontal branch morphology
 by \cite{RE04}
suggests an age difference of perhaps 4 Gyr between
the dominant metal-poor population and the most metal-rich stars.
\cite{HK04} reach a similar conclusion from their combined
spectroscopic and photometric study.  Indeed, it is now increasingly 
evident that the cluster contains
a number of distinct populations, each with their own kinematic and
chemical signatures (e.g.\ \cite[Norris et al.\ 1997,
Sollima et al.\ 2005]{NF97, EP03, SL05}).

We are also studying the abundance and age distribution of the
stars in this
cluster through the use of photometry and moderate resolution spectroscopy
of a large sample of cluster members on the main sequence and at the
cluster turnoff.  The results of that work will be presented elsewhere
(Stanford et al.\ 2005a, in prep).  In this contribution we concentrate
on what can be learned from the stars in our samples that do not follow
the relations exhibited by the majority of cluster members.  
In particular, among the bright red
giants in $\omega$ Cen there are a number of objects with anomalous
abundances, including stars with excesses of s-process elements
(e.g.\ \cite[Lloyd-Evans 1983, Vanture et al.\ 2002]{LE83, VW02}) 
and CH-stars with strong overabundances of carbon
(e.g.\ \cite[Bell \& Dickens 1974]{BD74}).  We investigate here the 
occurrence of similar anomalous stars on the main sequence, and discuss their 
possible origins.  A complete description of the results will be given in
Stanford et al.\ (2005b, in prep).

\section{Observations}

The stars observed were drawn from a photometric catalogue generated
from CCD images obtained with the 1m telescope at Siding 
Spring Observatory.  The catalogue contains $B$,$V$ photometry for ten
slightly overlapping 20$^{\prime}$ $\times$ 20$^{\prime}$ fields, centered
$\sim$20$^{\prime}$ from the cluster centre, that surround the cluster.
The particular targets for the spectroscopic observations consist of 
uncrowded (no
neighbours within 5$^{\prime\prime}$) stars in two magnitude ranges, a {\it
main sequence} sample with 18.0 $\leq$ $V$ $\leq$ 18.5 and
0.40 $\leq$ $B-V$ $\leq$ 1.10, and a {\it turnoff} sample with 
17.25 $\leq$ $V$ $\leq$ 18.0 and 0.60 $\leq$ $B-V$ $\leq$ 1.10.  
The candidates were observed with the
2dF multi-fibre instrument on the Anglo-Australian Telescope.  
The spectrograph setup employed gave useful 
coverage of $\sim$900\AA\/ centred at 4200\AA, with a resolution
of $\sim$2.5\AA\/ (FWHM).
 The availability of on-line data reduction 
software, and the high radial velocity of $\omega$ Cen (232 kms$^{-1}$)
meant that shortly after each observation, stars could be classified as either
cluster members or non-members.  Non-member stars were replaced with new 
candidates
and member stars were reobserved to build-up the signal-to-noise of their
spectra.  The final data set for the main sequence sample consists of 
spectra for just over 200 cluster 
members with typical S/N of $\sim$35 at the G-band. 

\section{Line Strength Indices}

\subsection{Metallicities}

The likely existence of an age-abundance relation in this cluster 
(e.g.\ \cite[Rey et al.\ 2004, Hilker et al.\ 2004]{RE04,HK04}) 
means that the determination of metal abundances
for the observed stars by simple techniques, such as a comparison of
line strengths with those of similar main sequence stars in clusters 
of known abundance, is not appropriate.  This will particularly be the
case for metal-rich $\omega$~Cen stars which are younger and 
therefore hotter, bluer and of weaker K-line strength than stars
of similar M$_{V}$ and abundance in standard clusters.
To overcome this situation we have
used the approach developed by \cite{TB99}, in which the derived abundance,
denoted by [Fe/H], is determined from appropriate combinations of the
equivalent width of the CaII K-line, the colour index $(B-V)_{0}$ and
the ACF index, which is the autocorrelation of the spectrum over the
region 4000--4280\AA\/ excluding the H$\delta$ line and the CN-bands in
the vicinity of 4215\AA.  Consideration of the errors in these quantities
indicates that the [Fe/H] values have a typical uncertainty of $\pm$0.2
dex.  
In the following we concentrate solely on the main sequence sample 
as it is relatively unbiased with respect to $(B-V)$ colour. 

\subsection{G-band (CH)}

\begin{figure}
\includegraphics[width=6cm,angle=-90]{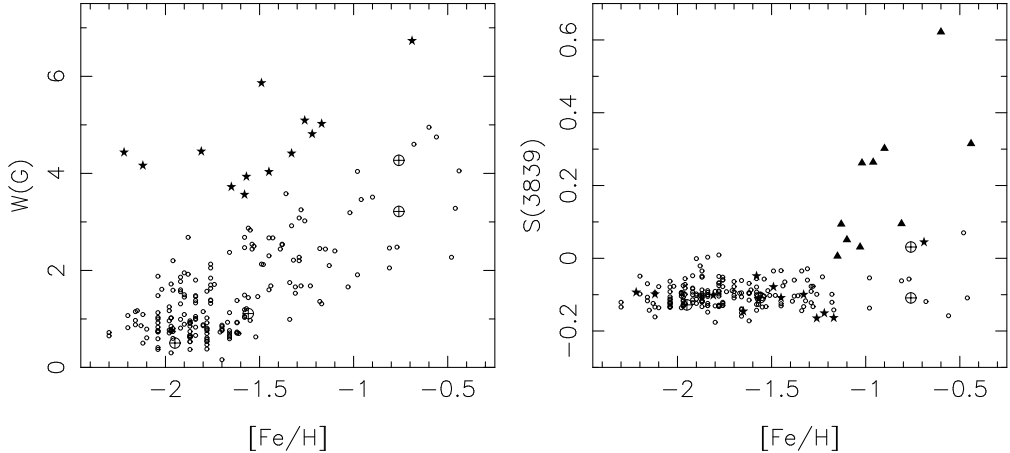}
\caption{(a) {\it Left panel:} The strength of the G-band index W(G) 
is plotted against
metallicity for the sample of $\omega$ Cen stars with V $>$ 18.
Stars with large values of W(G) for their abundance are plotted as
filled stars. (b) {\it Right panel:} The strength of the cyanogen index 
S(3839) is plotted against metallicity for same $\omega$ Cen sample. The 
strong G-band stars from the left panel are again plotted as filled stars,
while stars with large values of S(3839) for their abundance are plotted as
filled triangles.  In both panels 
the circled plus symbols are the mean locations for
similar samples of main sequence stars in the globular clusters
NGC~6397, NGC~6752 and 47~Tuc.  For the latter cluster there are
two points: the CN-strong stars are CH-weak and vice-versa. } \label{GBfig}
\end{figure}

To measure the strength of the CH features we have
computed the index W(G), which is a measure of the strength 
of the G-band (CH) at $\sim$4300\AA.  The values of W(G) are plotted
against [Fe/H] in the left panel of Fig.\ \ref{GBfig}.  Shown also
are the mean W(G) values for similar sets of main sequence
stars, observed with the same instrumental setup, in the standard globular
clusters NGC~6397 ([Fe/H]=--1.95), NGC~6752 (--1.56) and 47~Tuc 
\mbox{(--0.76)}.  
In 47~Tuc the main sequence
stars possess an anti-correlation between the strengths of CN and CH features,
and the stars can be classified as either CN-strong (CH-weak) or CN-weak 
(CH-strong).    
The two groups differ in carbon abundance with the CN-strong stars being 
depleted by $\sim$0.3 dex relative
to the CN-weak stars (cf.\ \cite[Cannon et al.\ 1998, Briley 
et al.\ 2004]{RC98,BH04}).  Consequently, there are two 47~Tuc points 
in the figure,
representing the two populations.  The $\omega$ Cen stars show a 
general increase in the W(G) values with increasing metallicity.  For the 
majority, the scatter in the W(G) 
indices  is largely driven by errors in the measured quantities,
particularly at the metal-poor end.  For the more metal-rich stars, however,
there are indications of an intrinsic spread in the W(G) values, with some
stars showing potentially significant C depletions similar to those
for the 47~Tuc CN-strong stars.  

Of particular interest are the 13 stars (plotted as filled
stars in Fig.\ \ref{GBfig}) that lie above the relation defined
by the majority of the stars -- these stars have larger W(G) values for
their [Fe/H] than a `typical' $\omega$ Cen main sequence member.  It
is immediately apparent from Fig.\ \ref{GBfig} that the `strong G-band' stars
are found at all metallicities: there is no indication that any
particular [Fe/H] is favoured over any other.  Given the
total sample size of 205 stars, the frequency of occurrence of these
strong G-band main sequence stars is of order 5\%, making them 
relatively rare objects.

To investigate the atmospheric abundances implied by the strong G-bands, 
we have carried out spectrum synthesis calculations.
The calculations assume our determination of [Fe/H] for each
star and use temperatures derived from the photometry.  In each case the
carbon abundance was varied until a satisfactory match was achieved between
the observed and calculated spectra.  We find that these strong G-band
stars are {\it typically enhanced in carbon by factors of $\sim$3 to 10}, 
occasionally more, relative to the `normal' $\omega$~Cen stars of comparable 
metallicity.  We also find that these stars
do not appear to require significant enhancements (or depletions) in N,
though we are sensitive to N abundance changes only for metallicities
above [Fe/H] $\approx$ --1.3 dex.  

\subsection{CN-bands}

In the right panel of Fig.\ \ref{GBfig} we show a plot of the cyanogen 
index S(3839) against
[Fe/H].  This index measures the strength of the CN bands at 3883\AA.
The panel also shows mean points for the three 
standard clusters, with the points for 47 Tuc again
showing the location of the CN-strong and CN-weak stars.  These two
47~Tuc 
populations, as well as differing in carbon abundances, 
differ in nitrogen abundances by a factor of $\sim$10 with the
CN-strong stars having the higher N abundances (\cite[Cannon 
et al.\ 1998, Briley et al.\ 2004]{RC98,BH04}).  The plot shows
that below [Fe/H] $\approx$ --1.3, we have little sensitivity to changes
in the CN-band strength.  This is a consequence of the moderate resolution of
the spectra, the temperature of the stars, and the fact that the CN-band
strength is proportional to the product of the C and N abundances, and
so decreases much faster with [Fe/H] than do CH (or NH) features.
Nevertheless, the plot reveals a number of interesting features.
First, the strong G-band stars identified in the left panel of Fig.\ 
\ref{GBfig} do not generally stand out from the bulk of the population in 
the right panel.  However, for [Fe/H] $\geq$ --1.3 where we have sensitivity
to CN-band strength changes, it is possible that some of the strong G-band 
stars are CN-weak, and that the most metal-rich example of these stars 
is CN-strong.  Second,
there is clearly a population of $\omega$ Cen main sequence stars which
do have strong, or in some cases extremely strong, CN-band strengths.  
These stars are plotted as filled triangles.  Of the 32
stars in our sample with [Fe/H] $\geq$ --1.3, 12, or approximately 40\%,
have strong cyanogen indices.  Spectrum synthesis calculations 
show that these CN-strong stars have {\it nitrogen enhancements of up to a
factor of $\sim$100} (the best fit for the star with the largest S(3839)
value has [C/Fe] $\approx$ +0.2 and [N/Fe] $\approx$ +2.0). 
In general, the best fits to the spectra of these CN-strong 
stars do not require large carbon depletions (or enhancements).

\subsection{SrII 4077}

\begin{figure}
\includegraphics[width=7.2cm,angle=-90]{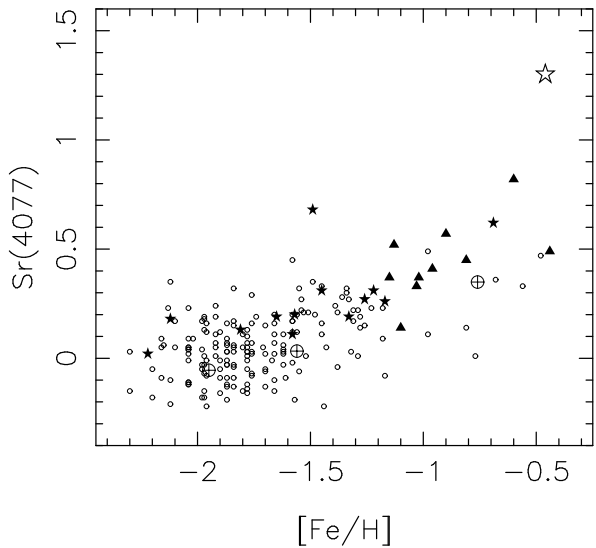}
\caption{The strength of the SrII line at 4077\AA\/ is plotted against
metallicity for the sample of $\omega$ Cen stars with V $>$ 18.
Symbols as for Fig.\ \ref{GBfig} except that the previously undistinguished
star with a large SrII strength is plotted as an open star.
The circled plus symbols are again the mean locations for
NGC~6397, NGC~6752 and 47~Tuc.  } \label{Srfig}
\end{figure}

In Fig.\ \ref{Srfig} we plot a measure of the strength of the SrII line
at 4077\AA\/ against [Fe/H] for our sample of $\omega$ Cen main sequence
stars.  Again the circled plus symbols represent
the mean locations for the standard clusters, noting that for 47~Tuc
there is no difference between the CN-strong and
CN-weak stars as regards SrII line strengths.  The $\omega$ Cen strong G-band
and strong-CN stars are plotted with the same symbols as for Fig.\
\ref{GBfig}.   Given the resolution and S/N of our spectra, the dispersion 
in the Sr(4077) values in Fig.\ \ref{Srfig} is largely
the result of observational uncertainties, particularly at lower abundances.
Equivalently, at this resolution and S/N, we are sensitive only to
large (positive) variations in the Sr abundance.  
We concentrate on three stars that
stand out in this figure, the strong G-band star at intermediate
metallicities, the CN-strong star at the high end of the metallicity
distribution, and a further metal-rich star that was previously 
undistinguished, i.e., its spectrum shows no evidence for significant C or N 
enhancements (or depletions).  Indeed spectrum synthesis calculations
for this star show that it may be mildly depleted in C ($\leq$0.3 dex)
and that it has an apparently normal N abundance.

The spectra of these 3 stars do indeed point
to substantial Sr overabundances.  In each case not only is the
SrII 4077\AA\/ line significantly enhanced but also the SrII line at
4215\AA\/ is unusually strong.  We have conducted spectrum synthesis 
calculations of the region surrounding the SrII 4077\AA\/ line for these 
stars and find that {\it overabundances of Sr of 
1.5 to 1.8 dex} are required to fit the line strengths.  We have also 
investigated the spectra of these Sr-strong stars in the vicinity 
of the BaII line
at 4554\AA.  The synthetic spectra calculations show that in both the 
(strong G-band, Sr-strong) star and the (CN-strong, Sr-strong)
star, the BaII line strengths are suggestive of
Ba overabundances comparable to that seen for strontium. However, for the star
without substantial C or N enhancements (or depletions), the BaII 4554\AA\/
line is barely visible, and Ba overabundances comparable to that seen
for Sr can be definitely ruled out.  

\section{Discussion}

As regards the C and N enhanced stars, it seems plausible
that we are seeing the effects of 3rd dredge-up combined with mass loss 
from intermediate-mass thermally-pulsing AGB (TP-AGB) stars.  The difference
between whether a star is C or N enhanced can then be ascribed to whether
or not hot-bottom burning occurred in the TP-AGB star to convert dredged-up
carbon into nitrogen.  What our data cannot answer, however,
is whether these enhanced abundances are a {\it surface effect}, e.g.\ 
resulting from mass transfer in a binary, or are {\it
intrinsic abundances}, i.e.\ the star formed from gas with these abundances.
The question is not easily tackled.  In the field, it is generally
accepted that the majority of CH-stars (and BaII-stars) are indeed binary
systems, with the observed atmospheric abundances resulting from wind-driven
mass transfer from the original primary to the companion (now the observed
star) during the primary's evolution through the TP-AGB phase 
(e.g.\ \cite[McClure 1984]{RM84}).  This is certainly also a reasonable
interpretation for the C-enhanced main sequence stars observed here, given
that they are found at all [Fe/H] values, that they have a  relatively low 
frequency of occurrence, which is likely to be compatible with a reasonable 
original total binary fraction, and that the relatively low central 
concentration of $\omega$ Cen means that the appropriate binaries are
not likely to have been disrupted before the original primary reaches the
TP-AGB phase.  

It is less obvious that this binary mass transfer explanation can also be 
applied for the N-rich stars.  While we have no constraints on the number 
of such stars for
abundances less than [Fe/H] $\approx$ --1.3, above this abundance the number
of N-rich stars is substantial ($\sim$40\% of the total).  Interpreting all
these stars as the result of mass transfer in binaries would seem to  
require an implausibly high original total binary fraction.  The alternative
is that these stars formed from material that was substantially
enhanced in nitrogen.  In this aspect, such a process may be related to that 
which drives the primordial abundance variations in other globular clusters,
as that phenomenon is known to occur in $\omega$ Cen (\cite[Norris \& 
Da~Costa 1995a]{ND95a}).  One potential method to investigate this
question is to compare the frequency of N-enhanced relatively
metal-rich giants with that observed in our main sequence sample.  If
the effect is primarily a mass-transfer driven atmospheric one, then
the frequency of occurrence in red giants should be reduced relative to
that in main sequence stars, as in at least some main sequence stars 
the current unusual surface abundances should be significantly diluted by
convective mixing as the star evolves up the red giant branch.  Conversely,
if the effect is primarily primordial, the main sequence and RGB 
frequencies should be comparable, although the possibility
that some stars could become N-enhanced through evolutionary mixing as they
ascend the 
RGB must also be considered.  The existing data for red giants do not provide 
any strong guidance. \cite{ND95b} list six stars with [Fe/H] $\geq$ --1.3
for which they have determined N abundances.  While noting the caveats
expressed in that paper concerning selection biases and potential systematic
errors in the N abundances, it is interesting that 3 of the 6 stars
are N-enhanced, nominally a similar ratio to that for the main sequence.
The size of the N-enhancements, however, typically a factor of 5--15
(\cite[Norris \& Da Costa 1995b]{ND95b}), are apparently notably less 
than for the main sequence stars.

As regards the Sr overabundances, those associated with C and/or N
enhancements can be ascribed to s-process nucleosynthesis occuring
in the same TP-AGB stars that provide the carbon and/or nitrogen enhancements.
This is particularly likely in those situations where Ba overabundances
are also seen.  What is much less clear is whether this mechanism could
also apply to the sole metal-rich star in our main sequence sample which has 
a large overabundance of Sr, but which lacks significant enhancements 
(or depletions) in C and N, and which does not have an overabundance of Ba 
matching that of strontium.  A high dispersion spectrum of this star, 
to allow determination of the abundances of a number of other 
elements (e.g.\ $\alpha$, O, Na, Al, Mg and r- and s-process), is  
required in order to understand its origin.

In summary, our moderate resolution spectra of a large sample of 
$\omega$ Cen main sequence stars have revealed a number of unusual objects.
Some of these stars may be the result of mass transfer in binary systems,
but others seem to require more exotic explanations.  Further observational
and theoretical studies are clearly necessary.



\end{document}